\documentclass[prl,aps,
twocolumn, showpacs,amsmath,amssymb]{revtex4}
\tighten
\usepackage{epsf}

\begin{document}
\title{Classical memory effects on  spin dynamics in two-dimensional systems}
\author{I.~S.~Lyubinskiy,  V.~Yu.~Kachorovskii}
\affiliation{A.F.~Ioffe Physical-Technical Institute, 26
Polytechnicheskaya street, Saint Petersburg, 194021, Russia}
\date{\today}
\begin{abstract}
{ We discuss classical  dynamics of electron spin in  two-dimensional semiconductors  with a
spin-split  spectrum. We focus on a special case, when spin-orbit induced random magnetic field is
directed along a fixed axis. This case is realized in III-V-based quantum wells grown in [110]
direction and also in [100]-grown quantum wells with  equal strength of Dresselhaus and  Bychkov-Rashba
spin-orbit couplings. We show that in such wells the long-time spin dynamics is determined by
non-Markovian memory effects. Due to these effects the non-exponential  tail $1/t^2$ appears in the
spin polarization.}
\end{abstract}
 \maketitle

Continuous reduction of the device sizes in last decades has initiated active research of
transport, optical and spin-dependent properties of low-dimensional nanostructures.
 In recent years, it was clearly understood  that not only quantum but also  purely classical
phenomena might lead to rich physics in such structures. In particular, a number of
non-trivial
transport phenomena, such as magnetic-field-induced classical localization \cite{fog,baskin},
high-field negative \cite{fog,baskin,bobylev,b4-1,curc0} and
positive \cite{mir1}  magnetoresistance, low-field anomalous magnetoresistance
\cite{an1,an2}, zero-frequency conductivity anomaly
\cite{pol}, and
non-Lorentzian shape of cyclotron resonance \cite{gor}
might be realized in  two-dimensional (2D) disordered systems. All these phenomena
arise due to classical non-Markovian memory effects which are neglected in the Drude-Boltzmann
approach. The strength
 of these effects is governed by a classical parameter
$d/l$ ($d$ is the characteristic scale of the disorder and $l$ is the transport scattering length).
Since the role of quantum effects is characterized by a parameter $\lambda/l$ ($\lambda$ is the
electron wavelength), the classical effects  might dominate in systems with long-range disorder,
where $d \gg \lambda.$


Usually, classical memory effects slow down relaxation processes leading to
non-exponential decay of correlation functions. In particular, the velocity autocorrelation
function in a 2D disordered system has a power tail \cite{ernst,hauge}
\begin{equation}
 \frac{\langle \mathbf v(t) \mathbf v(0) \rangle}{v_F^2}=-
 C~\frac{\tau^2}{t^2},~~~t\gg\tau,
 \label{corr}
\end{equation}
in contrast to exponential decay $\exp(-t/\tau)$ predicted by the Boltzmann equation.
 Here $\tau=l/v_F$
is the transport scattering time, $v_F$ is the Fermi velocity  and $C$ is the coefficient which
depends on the type of disorder: $C= 2d/3\pi l$ for the Lorentz gas model, where electrons scatter
on  hard disks of radius $d$ randomly distributed in a 2D plane with concentration $n$ ($nd^2\ll
1$) \cite{ernst}, and $C \sim d^2/l^2$ \cite{pol} for the scattering on the smooth random potential
with a characteristic scale $d$.
Physically, this long-lived tail
 is due to "non-Markovian memory" specific for diffusive returns to the same scattering
center \cite{hauge} (see also recent discussion in Refs.~ \cite{pol,remi}).

In spite of the large number of publications devoted to the study of non-Markovian transport
phenomena, the role of memory  effects in spin dynamics is not well understood. In this paper we
discuss the  slow down of the spin relaxation in 2D systems due to the non-Markovian memory. This
effect is of particular interest for new rapidly growing branch of semiconductor physics,
spintronics.  The main goal of spintronics is the development of novel electronic devices that
exploit the electron charge and spin on  equal footing \cite{avsh}.
 For effective functioning of such devices,
the  lifetime of the non-equilibrium spin must be  long compared to the device operation time.
In III-V-based semiconductor nanostructures, this requirement is  not easy to satisfy,
since in such structures the spin polarization relaxes rapidly due to Dyakonov-Perel (DP) spin relaxation mechanism
\cite{perel}. This mechanism predicts the exponential relaxation of non-equilibrium spin with
a certain
characteristic time $\tau_S.$
At small temperatures, this relaxation might slow down due to quantum interference effects
\cite{m2,my1}. However, with increasing temperature, the interference effects are suppressed by
inelastic scattering.
Here we show that the classical non-Markovian effects which are not very sensitive to the temperature
might  lead to long-lived $1/t^2$ non-exponential tail in the spin polarization in
analogy with velocity relaxation described by Eq.~\eqref{corr}.

The DP mechanism is based on the classical
picture of the angular spin diffusion in random magnetic field induced by spin-orbit coupling. In
2D systems, the corresponding spin-relaxation time $\tau_S$ is inversely proportional to the
momentum relaxation time $\tau$: $1/\tau_S \sim \langle\omega^2_{\mathbf p}\rangle \tau$
\cite{dyak}. Here $\omega_{\mathbf p}$ is the frequency of spin precession in a random magnetic
field, $\mathbf p$ is the electron momentum, and angular brackets denote averaging over momentum
directions (for $p \approx p_F$).
 As a consequence,  in high-mobility structures which are most promising for
device applications, $\tau_S$ is especially short. However,
 in some special cases, the relaxation of one of the spin
components can be rather slow even in a system with high mobility.
 In particular,  a number of recent researches
\cite{d1,d2,d3,d4,d5,d6} are devoted to GaAs symmetric quantum wells (QW) grown in $[110]$
direction. In such structures,  $\omega_{\mathbf p}$ is perpendicular to the QW plane \cite{dyak}
 and depends on one component of the in-plane momentum (say $x$-component)
\begin{equation}
\boldsymbol{\omega}_{\mathbf p}= \alpha p_x \hat {\mathbf z}. \label{omegaxi}
\end{equation}
Here $\hat {\mathbf z}$ is the unit vector normal to the well plane and $\alpha$ characterizes the
strength of the spin-orbit coupling. Also, the random magnetic field might be parallel to a fixed
axis  in an asymmetric $[100]$-grown QW due to the interplay between the bulk \cite{dress} and
structural \cite{rashba}   spin-orbit couplings \cite{pikus,golub,kim,loss1,loss2} (the structural
coupling depends on the gate voltage \cite{nitta}, so one can tune these two couplings to have
equal strength). For such QW Eq.~\eqref{omegaxi} is also valid, but in this case unit vector $\hat
{\mathbf z}$ is parallel to the QW plane.  In both cases, one component of the spin, $s_z$, does
not relax. Therefore, these structures are especially attractive for spintronics applications.

In this paper we discuss long-time dynamics of the spin polarization in such  structures. We
consider the relaxation of the vector $\mathbf s=(s_x,s_y)$ which is perpendicular to the random
magnetic field ($\mathbf s \perp\boldsymbol{\omega}_{\mathbf p}$) and show that similar to the
velocity autocorrelation function, the spin correlation function has long-lived $1/t^2$ tail
(both for the case of strong scatterers and for smooth potential). The analogy between velocity and
spin relaxation  is based
 on the following. As seen from Eq.~\eqref{omegaxi}, the spin
rotation angle is proportional to the integral $\varphi \sim \int p_x dt$ and is equal to zero for
 closed paths \cite{pikus}. Thus when electron returns to the impurity its spin restores the original direction. This implies
 some kind of memory
effects specific for the systems under discussion.

The Hamiltonian of the system  is given by
\begin{equation}
\hat H = \frac{\hat {\mathbf p}^2}{2m}+ \frac{\hbar}{2} \alpha p_x \hat{\sigma_z} + U(\mathbf r),
\label{hamilt}
\end{equation}
where $U(\mathbf r)$ is a random potential, $\hat \sigma_z$ is the Pauli matrix and $m$ is the
electron effective mass.

\begin{figure}
 \centerline{ {\epsfxsize=7.5cm{\epsfbox{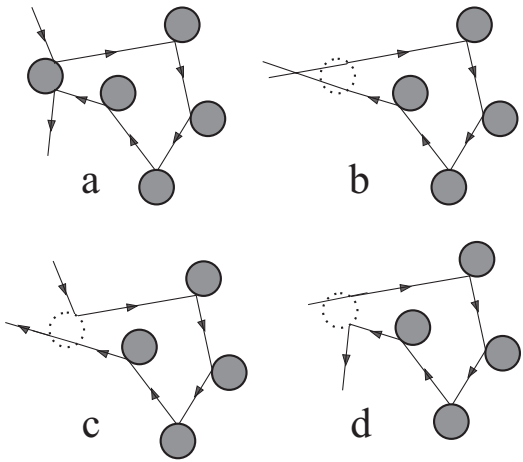}} }}
\vspace{-5mm}
\caption{ Four processes giving leading contribution to long-lived
tail in velocity and spin correlation functions. } \label{fig1}
\end{figure}

In the Boltzmann approach, the classical dynamics of spin related to Hamiltonian \eqref{hamilt}
is described by the kinetic equation \cite{perel}
\begin{equation}
\frac{\partial \mathbf s}{\partial t}+(\mathbf v \mathbf \nabla) \mathbf s = \hat J_B \mathbf s +
[\boldsymbol{\omega}_{\mathbf p}\times \mathbf s], \label{bol}
\end{equation}
where $\mathbf s(\mathbf r,\mathbf p)$ is the spin density related to the averaged spin as $\mathbf
 S   = \int \mathbf s(\mathbf r,\mathbf p) ~d^2\mathbf r d^2\mathbf p/(2\pi\hbar)^2$ and $\hat J_B$ is the
 Boltzmann
collision integral. Here we consider a case of degenerated electron gas ($T\ll E_F$), assuming that
the spin-polarized electrons have energies close to the Fermi energy $E_F$. First we assume
 that
electrons are scattered by strong scatterers randomly distributed in plane with average
concentration $n.$ In this case
\begin{equation}
\hat J_B \mathbf s(\theta)  = n v_F \int \sigma(\theta-\theta')[\mathbf s(\theta')-\mathbf
s(\theta))] d \theta', \label{JB}
\end{equation}
where $\sigma(\theta)$ is differential cross-section of one scatterer (for electrons with energy
$E\approx E_F$) and  we used shorthand notation $\mathbf s(\theta)= \mathbf s(\mathbf r,\mathbf p)$ ($\theta$ is
the angle of the vector $\mathbf p$). In Eq.~\eqref{JB} we neglected inelastic scattering. The role
of such scattering will be briefly discussed below. To account for classical memory effects we will
follow the method proposed in Refs.~\cite{kozub} (calculation of the velocity correlation function by this method
is presented in Ref.~\cite{remi}). The key idea is to replace   $ n \to
\sum_i \delta(\mathbf r - \mathbf r_i) = n+\nu(\mathbf r) $ in the collision integral, where $ \nu(\mathbf r) = \sum_i
\delta(\mathbf r - \mathbf r_i)-n, ~~\langle \nu(\mathbf r) \rangle = 0 $ (averaging is taken over
the position of the impurities). The collision integral becomes $\hat
J_B \to  \hat J_B + \hat J^*,$ where
\begin{align}
\hat J^*\mathbf s (\theta) = \nu(\mathbf r) v \int \sigma(\theta-\theta')[\mathbf
s(\theta')-\mathbf s(\theta)] d \theta'. \label{jnu}
\end{align}
By the following transformation
\begin{equation}
\mathbf s = \hat T(x)\mathbf s',\quad \hat T=\left[%
\begin{array}{cr}
  \cos qx & -\sin qx \\
  \sin qx &  \cos qx \\
\end{array}%
\right], \quad q=\alpha m
\label{transform}
\end{equation}
we eliminate the spin rotation term $[\boldsymbol{\omega}_{\mathbf p}\times \mathbf s]$ from
Eq.~\eqref{bol}
\begin{equation}
\frac{\partial \mathbf s'}{\partial t}+(\mathbf v \mathbf \nabla) \mathbf s' = (\hat J_B + \hat
J^*) \mathbf s' \label{tran1}
\end{equation}
(corresponding unitary transformation of Hamiltonian \eqref{hamilt} is presented in
Refs.~\cite{loss1, levitov}). Following \cite{remi}, we solve equation \eqref{tran1} treating the
term proportional to $\nu(\mathbf r)$ as a small correction. In the second order of perturbation
theory we obtain the following equation:
\begin{align}
&\frac{\partial \mathbf s'}{\partial t}+(\mathbf v \mathbf \nabla) \mathbf s' = \hat J_B \mathbf s'+
\delta \hat J \mathbf s', \label{kin}\\
 &\delta \hat J=\langle \hat J^* \hat G \hat J^* \rangle,
\label{ave}
\end{align}
where the kernel $G(\mathbf r, \varphi, \varphi', t)$ of the operator $\hat G$ obeys
\begin{equation}
\frac{\partial   G}{\partial t}+(\mathbf v  \nabla)
 G = \hat J_B  G +  \delta(\mathbf r)\delta(\varphi-\varphi')\delta(t).
\end{equation}
To calculate the average in the Eq.~\eqref{ave} we take into account that $\langle \nu(\mathbf
r)\nu(\mathbf r') \rangle = n \delta(\mathbf r-\mathbf r') $. As a result we get:
\begin{equation}
\delta \hat J \mathbf s'= v n \int_0^{\infty} dt' d\theta' ~ \delta \sigma(\theta-\theta', t')
[\mathbf s'(\theta', t-t')-\mathbf s' (\theta, t-t')],
\end{equation}
where
\begin{align}
&\delta \sigma(\theta-\theta', t)= v \int [\sigma(\theta-\varphi)-\sigma_0\delta(\theta-\varphi)]
\label{sig}  \\
& \times G(0, \varphi-\varphi', t)
[\sigma(\varphi'-\theta')-\sigma_0\delta(\varphi'-\theta')]d\varphi d\varphi' \nonumber
\end{align}
Here $\sigma_0=\int d\varphi\sigma(\varphi) $ is the total cross-section
 and
$G(0, \varphi-\varphi', t)=G(\mathbf r, \varphi,\varphi', t)|_{\mathbf r\to 0}$ is the probability
 for an electron starting in the direction $\mathbf n = (\cos\varphi, \sin\varphi)$ to return to the initial
 impurity after time $t$ along the direction $\mathbf n'=(\cos\varphi', \sin\varphi')$ (see Fig.~\ref{fig1}).
Four terms in the product $[\sigma(\theta-\varphi)-\sigma_0\delta(\theta-\varphi)]
[\sigma(\varphi'-\theta')-\sigma_0\delta(\varphi'-\theta')]$ correspond to four types of
correlations \cite{an2} shown in Fig.~\ref{fig1}. Fig.~\ref{fig1}a shows the process, where
electron experiences two real scatterings
 on the same  impurity. In Eq.~\eqref{sig} the corresponding contribution is
 presented by the term proportional to $\sigma(\theta-\varphi)\sigma(\varphi'-\theta').$
   In the process shown in Fig.~\ref{fig1}b
an electron  passes twice the region with the size of the order of impurity size without
scattering. Since the electron "keeps memory" about absence of impurity at a certain region of
space, there exists a correlation which is accounted for by the
term proportional to
$\sigma_0^2\delta(\theta-\varphi) \delta(\varphi'-\theta')$ in Eq.~\eqref{sig}. The interpretation
of the two other terms is based on the fact \cite{an2} that
 in the Boltzmann picture, which neglects
correlations, the following processes are allowed.  An electron scatters on an impurity and later
on passes through the region occupied by this impurity without a scattering (see
Fig.~\ref{fig1}c).
Another  process     is  shown in Fig.~\ref{fig1}d. The contributions of the terms
$\sigma_0\sigma(\theta-\varphi) \delta(\varphi'-\theta')$  and $\sigma_0\delta(\theta-\varphi)
\sigma(\varphi'-\theta')$  in Eq.~\eqref{sig} correct the Boltzmann result by substrating
probabilities of such unphysical events.

  At $t\gg\tau$
  the return probability reads
\begin{equation}
G(0, \varphi-\varphi', t)=\frac{1}{8\pi^2 D t}\left(1-\frac{l^2}{2Dt}~\mathbf n \mathbf n'\right)
 \label{wt0},
\end{equation}
where $l^{-1}=nv\sigma_{\rm tr}$ and $\sigma_{\rm tr}=\int d\varphi\sigma(\varphi)(1-\cos\varphi).$
Integrating Eq.~\eqref{wt0} over angles we get $1/4\pi D t$ which is the probability to return to
the initial point with arbitrary angle (diffusive return). The second term in  Eq.~\eqref{wt0} is a
small angle-dependent correction which is responsible for the  effect under discussion.
Indeed, one can see that the  first term in Eq.~\eqref{wt0} gives zero contribution to
Eq.~\eqref{sig}. Calculating the contribution of the second term we get
\begin{equation}
\delta\sigma(\theta,t)=-\frac{\sigma_{tr}^2}{4\pi^2 v} \frac{\cos\theta}{t^2}. \label{correc}
\end{equation}
It worth noting that $\delta \sigma(\theta,t)dt$ has a dimension of length and can be interpreted
as a correction to the scattering cross-section due to diffusive returns taking the  time lying in
the interval  $[t, t+dt].$

The diffusion equation can be  obtained by standard means     from kinetic equation \eqref{kin}
with the
use of Eq.~\eqref{transform}. As a result we find the diffusion-like equation for the isotropic
part of the spin density $s_0=\langle s(\mathbf p,\mathbf r,t)\rangle_{\theta}$ (averaging is taken
over momentum directions):
\begin{equation}
\frac{\partial \mathbf s_0}{\partial t}= D \tilde\Delta \left(\mathbf s_0 -
\frac{n\sigma_{tr}^2\tau}{4\pi} \int_{\tau}^{\infty} \frac{ \mathbf s_0(t-t')dt'}{t'^2}\right),
\label{fin}
\end{equation}
where $\tilde{\Delta} = \hat T(x) \Delta \hat T(x)^{-1} = (\partial/\partial x+q\hat
\epsilon)^2+\partial^2/\partial y^2$ and $\hat \epsilon$ is the antisymmetric tensor:
 $\hat \epsilon \mathbf s = [\mathbf e_z\times\mathbf s]$. Eq.~\eqref{fin} describes the spin dynamics in the diffusion
 approximation. It simplifies in the homogeneous case:
\begin{equation}
 \frac{\partial\mathbf s_{0}}{\partial t} = - \frac{ \mathbf s_{0}}{\tau_S} +
\frac{n\sigma_{tr}^2\tau}{4\pi\tau_S} \int_{\tau}^{\infty} \frac{ \mathbf s_0(t-t')dt'}{t'^2}.
\label{boltz1}
\end{equation}
Here $1/\tau_S= Dq^2= (\alpha p_F)^2\tau/2$
is the Dyakonov-Perel' spin relaxation rate. The initial
condition for \eqref{boltz1} is $\mathbf s_0(0)=s_{\rm i}$   (we also assume that
 $ \mathbf s_0(t)=0$ for $t<0$). Neglecting the
second term in the  rhs of Eq.~\eqref{boltz1} we get the exponential relaxation $\mathbf
s_{0}(t)=\exp(- t/\tau_S)\mathbf s_{\rm i}$ \cite{perel}. This solution is valid until $\exp(-
t/\tau_S) \sim n \sigma^2_{tr}\tau \tau_S/t^2.$ For larger times, $  \int^{t}_{\tau} dt' \mathbf
s_{0}(t-t')/t'^2 \approx \mathbf s_{\rm i}\tau_S/t^2$ and one can
 neglect the term
$\partial\mathbf s_0/\partial t$ in Eq.~\eqref{boltz1}. As a result   we find that the  polarization has a
long-lived  tail
\begin{equation}
\mathbf s_{0}(t) \approx \frac{n\sigma_{tr}^2 \tau\tau_S}{4\pi t^2} \mathbf s_{\rm i}=
\frac{\sigma_{tr}}{l} \frac{\tau\tau_S}{4\pi t^2} \mathbf s_{\rm i}, \label{tail}
\end{equation}
which is positive  in contrast to  Eq.~\eqref{corr}.

 Eq.~\eqref{tail} was derived for the case of strong scatterers with low concentration ($nd^2
\ll 1$). The opposite limiting case (weak scatterers, $nd^2 \gg 1$) corresponds to the smooth
random potential with the correlation function
$\langle U(\mathbf r)U(\mathbf r') \rangle = \int\kappa_q \exp[i\mathbf
q(\mathbf r-\mathbf r')] d^2\mathbf q/(2\pi)^2$. In this case, the collision integral
can be written as a sum of the
Boltzmann
collision integral $\hat J_B = (1/\tau)(\partial^2/\partial\varphi^2)$ and
\begin{equation}
\label{Jsmth} \hat J^* =\frac{\partial}{\partial\varphi} \left[ \int_0^\infty dt [\mathbf n \times
\mathbf f(\mathbf r ) ] [\mathbf n \times \mathbf f(\mathbf r- \mathbf v t) ] - \frac{1}{\tau}
\right]\frac{\partial}{\partial\varphi},
\end{equation}
where
 $1/\tau=(1/2\pi m^2 v_F^3)\int_0^{\infty} q^2
\kappa_q dq$ and  $\mathbf f = - \nabla U(\mathbf r)/mv_F$ \cite{d/dr}. One can check that $\langle \hat J^* \rangle = 0.$ Substituting
Eq.~\eqref{Jsmth} into Eq.~\eqref{ave}, using Eq.~\eqref{wt0} and accounting for two types of
correlations \cite{pol} we get
\begin{align}
&\delta \hat J \mathbf s(\varphi) =\frac{2\pi^2d^2}{\tau^2} \int_0^{\infty}dt' [ G''(0,t')
\frac{\partial^2 \mathbf s(\varphi,t-t')}{\partial \varphi^2}~+ \nonumber
\\
&G''(\pi,t') \frac{\partial^2 \mathbf s(\varphi+\pi,t-t')}{\partial \varphi^2} ] ,\label{jsmooth}
\end{align}
where
 $G''(\varphi,t) =
\partial^2 G(0,\varphi,t) / \partial \varphi^2.$   The calculations analogous to
the case of strong scattering centers yield
\begin{equation}
\mathbf s_0(t)= \frac{d^2}{l^2}\frac{\tau \tau_S}{t^2}~ \mathbf s_{\rm i}. \label{ssmooth}
\end{equation}
In Eqs.~\eqref{jsmooth} and \eqref{ssmooth},
$d=\sqrt{2\int_0^{\infty}\kappa_q^2q^3dq}/\int_0^{\infty}\kappa_q q^2 dq.$

Above  we assumed that $\mathbf s_{\rm i}$ is homogenous. For slowly varying $\mathbf s_{\rm
i}(\mathbf r),$ the derived equations relate  $\mathbf s_0(\mathbf r,t)$ with $\mathbf s_{\rm
i}(\mathbf r)$ provided that the spatial scale of inhomogeneity $L$ is large compared to
$\sqrt{D\tau_S}\sim 1/m\alpha$ \cite{D/G}. One can show that in the opposite case $ l \ll L \ll
1/m\alpha,$ these equations are also valid relating  $\int d\mathbf r ~\mathbf s_{0}(\mathbf r,t)$ with
$\int d\mathbf r ~\mathbf s_{\rm i} (\mathbf r).$

Let us briefly discuss the role of electron-electron interaction. Such interaction manifests itself
both in inelastic scattering and in the additional, with respect to the diffusion process, decay of
density fluctuation due to Maxwell relaxation. Such a relaxation partially suppresses the
long-lived tail in velocity correlation function \eqref{corr}
 leading to the faster  decay \cite{ufn}: $1/t^2 \to r^2_B/Dt^3,$ where $r_B$ is the
 2D screening length, which coincides with the
Bohr radius. In contrast to this, the Maxwell relaxation has no  effect on the spin dynamics because
in the classical approximation the spin fluctuations are not coupled to the charge fluctuations and,
as a consequence, do not lead to creation of long-range
electrical field responsible
 for Maxwell relaxation.

As for  electron-electron  collisions, their characteristic time
$\tau_{ee}$ is inversely proportional to $T^2.$ For relatively small temperatures, $\tau_{ee} \gg \tau$ and
electron-electron collisions do not
have any effect on the  spin relaxation in the classical approximation. In the opposite limiting case,
$\tau_{ee} \ll \tau,$ electron-electron collisions might suppress  spin relaxation \cite{glazov}.
The detailed discussion
of this case is out of the scope of this paper. We believe that $1/t^2$ dependence of long-time polarization
 is also valid  for this case,  while the coefficient in this dependence might change.

Finally, we compare non-Markovian tail in the spin polarization (Eqs.~\eqref{tail},
\eqref{ssmooth}) with the long-lived tail induced by weak localization \cite{my1}: $ \mathbf
s_0(t)=\mathbf s_{\rm i}(\tau_S/\pi k_F l t)  \exp(-t/\tau_{\varphi}),$ where $\tau_{\varphi}$ is
the phase-breaking time.  At $T=0,$ when $\tau_{\varphi}=\infty,$ long-time spin dynamics is
determined by weak localization. However, for $T\neq 0$
classical memory
effects dominate at $t\gg\tau_{\varphi}.$

In conclusion, we developed a theory of long-time spin dynamics for 2D system, where
spin-orbit-induced magnetic field is parallel to a fixed axis. We showed that independently on the
type of disorder  the non-equilibrium spin polarization in such a system decays as $1/t^2$ (for
$t\to \infty$) due to purely classical memory effects.
\begin{acknowledgments}
 This work has been supported by  RFBR, a grant of the
RAS, a grant of the Russian Scientific School, and a grant of the foundation "Dynasty"-ICFPM.
\end{acknowledgments}


\begin{thebibliography}{}
\bibitem{cross}
In a purely classical case $\sigma_0=\infty$ (with an exception for  Lorentz gas model) due to
divergent contribution of small-angle scattering.
However,  our final result only depends on $\sigma_{\rm tr}$
(see Eq.~\eqref{correc}) thus justifying the  approach based on classical consideration.
\bibitem{d/dr}
Actually,
collision integral
also contains
terms with spatial derivatives.
However,
they give negligible contribution to
velocity and
spin correlation functions.
\bibitem{D/G} Note that $1/m\alpha$ is a characteristic scale of
non-decaying spin oscillations which might exist in the system
under discussion (see  Ref.~\cite{loss1}).
\end{thebibliography}
\end{document}